\documentclass[conference]{IEEEtran}
\IEEEoverridecommandlockouts
\usepackage{cite}
\usepackage{bbold}
\usepackage{amsmath,amssymb,amsfonts}
\usepackage{bbm}
\usepackage{algpseudocode}
\usepackage[multiple]{footmisc}
\usepackage{algorithm}
\usepackage{graphicx}
\usepackage{textcomp}
\usepackage{multirow}
\usepackage[dvipsnames]{xcolor}
\usepackage{balance}
\usepackage{booktabs}
\usepackage{listings}
\usepackage{threeparttable}
\usepackage{wrapfig}
\usepackage{todonotes}
\def\BibTeX{{\rm B\kern-.05em{\sc i\kern-.025em b}\kern-.08em
    T\kern-.1667em\lower.7ex\hbox{E}\kern-.125emX}}
\usepackage{eso-pic}
    
\usepackage[hyperfootnotes=false]{hyperref}

\newcommand\copyrighttext{%
  \footnotesize \textcopyright 2023 IEEE. Personal use of this material is permitted.
  Permission from IEEE must be obtained for all other uses, in any current or future
  media, including reprinting/republishing this material for advertising or promotional
  purposes, creating new collective works, for resale or redistribution to servers or
  lists, or reuse of any copyrighted component of this work in other works.
  DOI: \href{https://doi.org/10.1109/IPCCC59175.2023.10253884}{https://doi.org/10.1109/IPCCC59175.2023.10253884}}
\newcommand\copyrightnotice{%
\AddToShipoutPicture*{%
\put(45,30){%
\centering
\fbox{\parbox{\dimexpr\textwidth-\fboxsep-\fboxrule\relax}{\copyrighttext}}
}}}

\DeclareMathAlphabet{\altmathcal}{OMS}{cmsy}{m}{n}

\colorlet{punct}{red!60!black}
\definecolor{background}{HTML}{EEEEEE}
\definecolor{delim}{RGB}{20,105,176}

\def\myvariablecolor{black}

\lstdefinelanguage{json}{
basicstyle=\ttfamily\footnotesize,
numbers=none,
numberstyle=\scriptsize,
stepnumber=1,
captionpos=b,
keepspaces=true, 
numbersep=5pt,
showspaces=false,
showstringspaces=false,
showtabs=false, 
tabsize=2,
breaklines=true,
morecomment=[l][\color{red}]{\#},
backgroundcolor=\color{background},
literate=
  {:}{{{\color{punct}{:}}}}{1}
  {,}{{{\color{punct}{,}}}}{1}
  {\{}{{{\color{delim}{\{}}}}{1}
  {\}}{{{\color{delim}{\}}}}}{1}
  {[}{{{\color{delim}{[}}}}{1}
  {]}{{{\color{delim}{]}}}}{1},
}

\begin{document}

\title{Karasu: A Collaborative Approach to Efficient Cluster Configuration for Big Data Analytics}

\author{
\IEEEauthorblockN{
Dominik Scheinert\IEEEauthorrefmark{1},
Philipp Wiesner\IEEEauthorrefmark{1},
Thorsten Wittkopp\IEEEauthorrefmark{1},
Lauritz Thamsen\IEEEauthorrefmark{2},
Jonathan Will\IEEEauthorrefmark{1},
and Odej Kao\IEEEauthorrefmark{1}
}
\IEEEauthorblockA{
\IEEEauthorrefmark{1}
Technische Universit{\"a}t Berlin, Germany, \{firstname.lastname\}@tu-berlin.de}
\IEEEauthorblockA{
\IEEEauthorrefmark{2}
University of Glasgow, United Kingdom, lauritz.thamsen@glasgow.ac.uk}
}

\maketitle
\copyrightnotice

\begin{abstract}
Selecting the right resources for big data analytics jobs is hard because of the wide variety of configuration options like machine type and cluster size.
As poor choices can have a significant impact on resource efficiency, cost, and energy usage, automated approaches are gaining popularity.
Most existing methods rely on profiling recurring workloads to find near-optimal solutions over time.
Due to the cold-start problem, this often leads to lengthy and costly profiling phases.
However, big data analytics jobs across users can share many common properties: they often operate on similar infrastructure, using similar algorithms implemented in similar frameworks.
The potential in sharing aggregated profiling runs to collaboratively address the cold start problem is largely unexplored.

We present \emph{Karasu}, an approach to more efficient resource configuration profiling that promotes data sharing among users working with similar infrastructures, frameworks, algorithms, or datasets.
Karasu trains lightweight performance models using aggregated runtime information of collaborators and combines them into an ensemble method to exploit inherent knowledge of the configuration search space.
Moreover, Karasu allows the optimization of multiple objectives %
simultaneously.
Our evaluation is based on performance data from diverse workload executions in a public cloud environment.
We show that Karasu is able to significantly boost existing methods in terms of performance, search time, and cost, even when few comparable profiling runs are available that share only partial common characteristics with the target job.

\end{abstract}


\section{Introduction}
\label{sec:introduction}

Reliable, timely, and efficient data processing is of highest importance as the amount of data in industry and research continues to grow.
However, the optimal selection and configuration of resources for user-specific goals and constraints remains challenging~\cite{Google2020Autopilot, cloudcomputingchallenges2018,RajanKCK16}.
To assist users, various approaches have been proposed over time, which can be grouped into two categories:
Model-based methods often require access to historical performance data~\cite{venkataraman2016ernest,RajanKCK16,ShahAKW19,AlSayehS19,KirchoffXMR19,ChenLLWZ21silhouette,AlSayehMJPS22} to train complex performance models, but data on past workload executions is not always available.
In contrast, search-based methods~\cite{AlipourfardLCVY17,HsuNFM18,hsu2018micky,bilal2020finding,klimovic2018selecta,fekry2020accelerating,MendesCRG20,LiuXL20,SongZLSFDS21,CasimiroD0RZG20} perform profiling runs using all or a subset of the input data to iteratively build performance models and to converge on an optimized resource configuration for a (recurring) workload.
These methods often incorporate some form of Bayesian Optimization (BO) for deciding which resource configurations to profile next.
Since past execution data is often not available and each workload has a unique execution behavior that must be understood, search-based methods are more practical because they always work for the situation at hand~\cite{0007SCR20}.

Unfortunately, in practice, methods from both categories are often not applicable to users with infrequent and varying processing needs such as scientists or users in small and medium-sized businesses.
For these users, performance data availability is often limited and the cold-start problem introduces significant overhead.
However, data analysis jobs of different users often show a high degree of similarity:
jobs are often executed on the same infrastructure (like public clouds), implementing popular algorithms (like K-means clustering or principal component analysis), on common frameworks (like Hadoop or Spark), and even datasets.
Thus, we expect a considerable potential for sharing profiling information among users to build more sophisticated and capable performance prediction models.
In recent related~\cite{fekry2020accelerating,FekryCPRH20} as well as own~\cite{ScheinertTZWAWK21,will2021c3o} work, ideas are presented that generally qualify for a collaborative scenario.
However, to allow as many collaborators as possible to engage in such efforts, it is important to implement sharing in a way that does not require the disclosure of detailed and potentially sensitive runtime information.

We propose \emph{Karasu}, a novel approach to enhance existing search-based methods for more efficient resource configuration profiling.
Unlike previous search-based approaches, we explicitly consider a collaborative scenario:
By utilizing models fitted on aggregated profiling data of collaborators, Karasu obtains a better understanding of the potentially large resource configuration search space via ensemble learning.
This allows for informed decisions and thus less profiling runs to find near-optimal configurations.
Moreover, Karasu directly supports the simultaneous optimization of multiple objectives such as runtime, cost, or energy usage.

\emph{Contributions}. The contributions of this paper are:

\begin{itemize}
    \item An approach to resource configuration profiling that extends existing methods by incorporating shared historical performance data.
    It is inherently designed to support the integration of multiple criteria during optimization.
    \item A method for aggregating low-level and potentially sensitive resource and performance metrics in a way that enables the similarity-based exchange of knowledge artifacts while retaining workload-specific details.
    \item An openly available implementation\footnote{\url{https://github.com/dos-group/karasu-cloud-configuration-profiling}} of Karasu which was evaluated with regard to various relevant metrics. 
\end{itemize}

\section{Resource Configuration Profiling}
\label{sec:rc_profiling}
In this section, we formalize the problem of resource configuration profiling, highlight common challenges, and discuss the assumptions specific to our contribution.

\subsection{Problem Formalization}
Machine types are a commonly used abstraction for bundles of resources, whether in private cluster environments or public cloud environments.
We define a resource \emph{configuration} $c_j \in C$ to be constituted by a machine type and a machine count.
On the other hand, a processing \emph{workload} $w_i \in W$ is characterized by the functionality and version of the \emph{framework} with which it was realized, the \emph{algorithm} implementation and parametrization used, and the size and characteristics of the \emph{dataset} to be processed.  
When a workload $w_i$ is run with a particular resource configuration $c_j$, both $n$ metrics (e.g. CPU or memory utilization) over time and machines, referred to as $l_{ij} \in \mathbb{R}^{n \times t}$, as well as $m$ final performance measures (e.g. cost, runtime) denoted as $y_{ij} \in \mathbb{R}^m$ can be obtained and recorded, where $t$ represents the number of individual metric values over time and across machines.
More specifically, these noisy performance measures are obtained by running a workload with a specific resource configuration in the target environment, where we attempt to iteratively approximate this black-box function via a model $f$.
The model then attempts to learn the resource-dependent behavior of a workload by estimating the mapping of profiled resource configurations to associated performance measures, while minimizing formulated objectives and satisfying defined constraints.
Moreover, when attempting to find a near-optimal resource configuration, it is also often the case that the search phase itself is constrained with respect to the performance measures of interest, e.g. when only a limited financial budget is available to cover the cost overhead of profiling in a public cloud environment.

In this paper, we assume that certain objectives -- in our evaluation \emph{cost} and \emph{energy consumption} -- shall be minimized while resource configuration candidates are also required to obey to \emph{runtime constraints} of a workload. 
Hence, we consider a constrained optimization problem.

\begin{figure*}%
    \centering
    \includegraphics[width=1\textwidth]{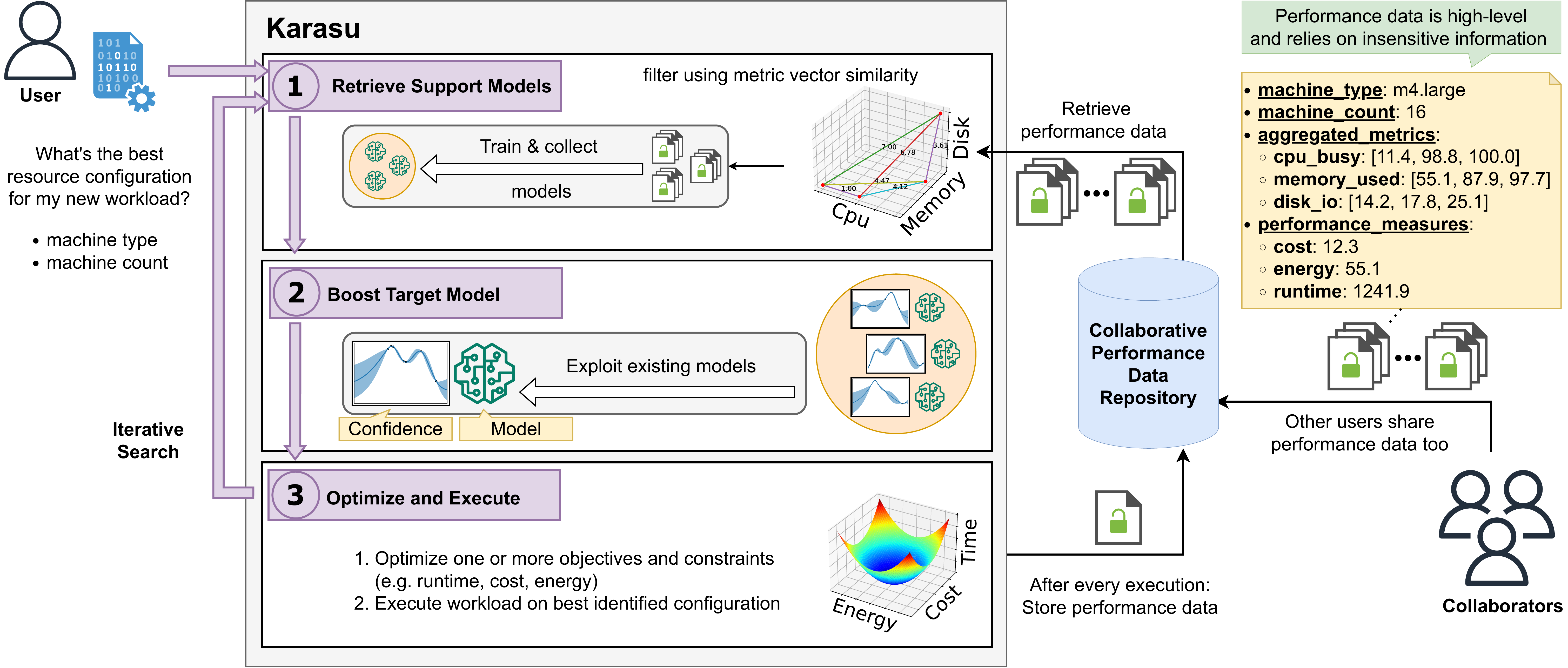}
	\caption{In each iteration, Karasu employs lightweight performance models from a repository using a similarity-based selection strategy operating on metric vectors.
	These models are used to improve the local performance model of a particular target workload following an ensemble approach.
	Under consideration of formulated objectives and constraints, the workload is then executed with the most promising configuration.
	}
	\label{fig:overview}
\end{figure*}

\subsection{Challenges \& Assumptions}
\label{sec:challengesassumptions}
From the introduced problem formalization of profiling suitable resource configurations for a target workload, we can deduce multiple downstream challenges and assumptions.

\textbf{Challenges.} Historical data on workload executions for performance modeling is often unavailable in practice, resulting in the \emph{cold-start problem} and usually requiring a costly and lengthy initial profiling phase, which is not ideal for many practitioners. 
Although prior works of ours~\cite{ScheinertTZWAWK21,will2021c3o} motivate collaborative approaches that involve data sharing, they strongly focus on common algorithm implementations from established frameworks and pay less attention to data confidentiality concerns, i.e., \emph{data minimalism} during the sharing phase is not actively pursued, which can hinder people to collaborate.
Finally, the majority of existing methods are most concerned with finding resource configurations that minimize cost and, if necessary, adhere to defined runtime targets of a workload -- they typically do not support \emph{optimization for multiple goals at once}, which is becoming more important as, for example, increasingly also the carbon footprint of applications needs to be optimized~\cite{WorldBank_CarbonPricing_2020, Eilam_CloudCarbonAccounting_2021}.

\textbf{Assumptions.} Our goal is to address these challenges and overcome the limitations of related work.
Since our approach is intended to extend existing solutions, we rely on similar assumptions.
We assume \emph{homogeneous cluster configurations}, which is reasonable since 1) we are primarily considering runtime-constrained workloads and thus batch processing applications with a finite and often constant amount of data to process, 2) modeling the effects of elastic scaling of a workload quickly becomes non-trivial and requires more insightful data, which conflicts with data minimalism considerations, and 3) established frameworks usually assume a set of homogeneous resources by design.
Also, as in related work, we assume that workloads are \emph{executed on single-purpose clusters} and do not initially compete with other data-intensive workloads of the same user concurrently, which ensures fairly informative but noisy recordings of metrics and performance measures that can be handled with methods such as BO.

\section{Approach}
\label{sec:approach}
This section presents the main ideas behind \emph{Karasu} and how it can be used to collaboratively accelerate resource configuration profiling for data processing workloads.
An overview of the approach is depicted in \autoref{fig:overview}.

\subsection{Overview}

Karasu improves existing approaches for resource configuration profiling through the use of aggregated profiling runs shared by collaborators.
As illustrated in~\autoref{fig:overview}, it consists of three steps and associated components, which are described in detail in the following sections:
\begin{enumerate}
    \item An opportunistic selection strategy based on workload resource requirements that allows the selection of promising BO candidate models from shared repositories without requiring detailed sensitive workload information (\autoref{sec:approach_selecting}). 
    \item An internally employed ensemble method for Gaussian Processes, which leverages the lightweight BO models of other workloads to accelerate the resource configuration search for a new target workload (\autoref{sec:approach_sharing}).
    \item A generalization of our approach that supports multi-objective optimization (\autoref{sec:approach_optimizing}).
\end{enumerate}
Note that hereafter, we present these chronological steps in a slightly different order to better connect the previously introduced problem formulation with our approach.

\subsection{Sharing and Leveraging Data and Models}
\label{sec:approach_sharing}

By our previously introduced definitions, running a workload $w_i$ with a resource configuration $c_j$ can be summarized after execution by the tuple $(w_i, c_j, l_{ij}, y_{ij})$.
Related work has shown that normal BO can already work with a subset of this information by learning a mapping from resource configurations to observed performance measures~\cite{AlipourfardLCVY17}.
In other words, the behavior of a workload in a target environment can be modeled using only this information, i.e., there is no need to encode further details such as libraries, dependencies, or workflow structures in the model input.
By building a probabilistic model $f(\mathbf{x}|D)$ from observations $D$ for a single objective only, i.e. 
\begin{equation*}
    D = \{(\mathbf{x}_j, y_j)| \forall c_j \in C: \mathbf{x}_j = h(c_j), y_j = y_{ij}(k)\}
\end{equation*}
where $\mathbf{x}_j$ denotes the resource configuration encoding produced by an encoder function $h$ and $y_j$ the desired $k$-th performance measure used as target variable, the model used to approximate the black-box function can be sufficiently fit after a handful of profiling runs. 
As shown in related work~\cite{AlipourfardLCVY17,HsuNFM18}, $h$ can be designed to deterministically encode the properties of a resource configuration, e.g. interconnect speed or GPU availability of machines, into a vector in a sufficient and ideally discretized way.
The latter significantly reduces the search space of candidate resource configurations. 
Finally, the generated multidimensional encodings are used as model input, and the bounds and specifications of the encoder function are used to describe the search space.
Now, to build a model based on historical data, that data must either \raisebox{.5pt}{\textcircled{\raisebox{-.9pt} {1}}} come from the same workload, or \raisebox{.5pt}{\textcircled{\raisebox{-.9pt} {2}}} include more discriminative features (e.g. workload parameters, software versions) that can be used as additional model input for discrimination and improved learning.
In a scenario where users share performance data, the former is an unrealistic assumption, while the latter most likely requires the sharing of sensitive information, which we want to minimize. 

\begin{figure}[b!]
    \centering
    \includegraphics[width=1\columnwidth]{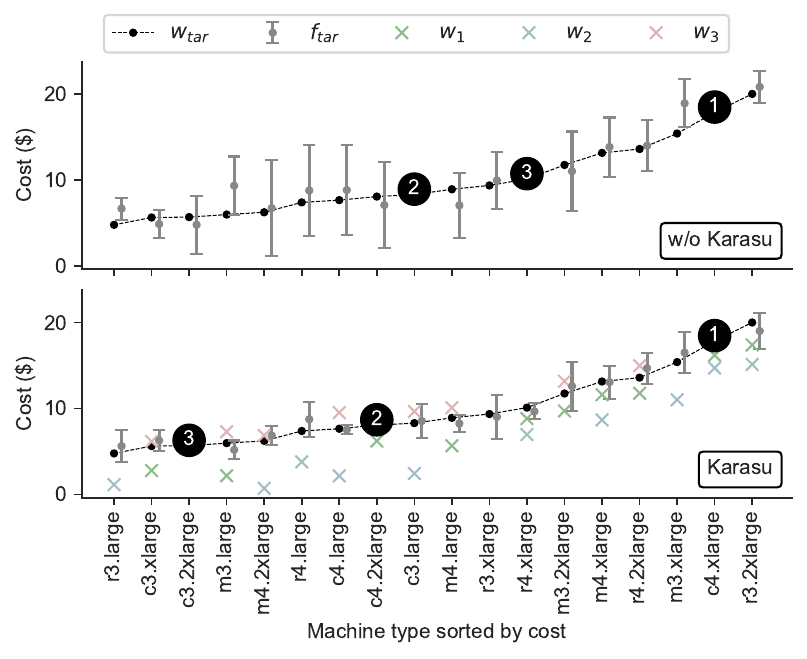}
    \caption{ 
    \textbf{Top:} Using only the data obtained from profiled resource configurations (enumerated circles) for $w_{tar}$ leads to an uncertain and slowly learning model, which initially relies also more on random sampling. 
    \textbf{Bottom:} With Karasu, we can make use of the profiling information from $w_1, w_2, w_3$ to improve profiling decisions by estimating an ensemble function that faster localizes the desired optimum for $w_{tar}$ and requires less random sampling.}
    \label{fig:karasu_intuition}
\end{figure}

\paragraph{Leveraging} 
We propose to build a BO model $f_i$ (Gaussian Process) for each individual workload $w_i \in W$ using its associated data, and then use a (sub)set of these workloads ($W' \subseteq W$) and their associated models to estimate a new model $f_{tar}$ for a target workload $w_{tar}$.
We do this by borrowing from the general idea of ranking-weighted Gaussian process ensemble (RGPE) explained in~\cite{feurer2022practical}:
\begin{equation*}
    f_{tar}(\mathbf{x}|D_{tar}) = \sum_{w_i \in W'} a_i f_i(\mathbf{x}|D_i).
\end{equation*}
Here, $a_i$ denotes the weight associated with the workload $w_i$ in a given ensemble learning procedure, and $D_i$ the corresponding observations. 
Conveniently, $f_{tar}$ remains a model built by a Gaussian Process, which allows access to a wide range of existing methods developed for BO:
\begin{equation*}
f_{tar}(\mathbf{x}|D_{tar}) \sim \altmathcal{N}\bigg( \sum_{w_i \in W'} a_i \mu_i(\mathbf{x}), \sum_{w_i \in W'} a_i^2 \sigma_i^2(\mathbf{x}) \bigg)
\end{equation*}
Here, $\mu_i$ and $\sigma_i^2$ denote the mean and variance parametrizing the BO model for a workload $w_i$. 
To weight the different models appropriately, RGPE first estimates the capabilities of each individual model $f_i$ using the number of misranked pairs as loss, i.e.
\begin{equation*}
    \mathcal{L}(f_i, D_i) = \sum_n^{|D_i|} \sum_m^{|D_i|} \mathbb{1} ((f_i(\mathbf{x}_n) < f_i(\mathbf{x}_m)) \oplus (y_n < y_m)),
\end{equation*}
where $\oplus$ denotes the exclusive-or operator.
In a second step, a fixed number of samples $s$ are drawn from the posterior of each individual model $f_i$ and evaluated using the above ranking loss, i.e. $\mathcal{\ell}_{i, s'} \sim \mathcal{L}(f_i, D_i)$ for $s'=1, \ldots, s$, such that a weight $a_i$ for $f_i$ is computed as follows:
\begin{equation*}
   a_i = \frac{1}{s} \sum_{s'=1}^s \left( \frac{\mathbbm{I}(i \in \arg \min_{i'} \mathcal{\ell}_{i', s'})}{\sum_{w_j\in W'} \mathbbm{I}(j \in \arg \min_{i'} \mathcal{\ell}_{i', s'})} \right).
\end{equation*}
For Karasu, this design has several advantages, such as the reuse of existing trained models and thus the mitigation of the cold-start problem, since already established (partial) knowledge about the resource configuration search space can be used.
This is possible because, for the ranking loss of RGPE, the actual prediction values do not matter for the optimization, since only the desired optimum needs to be localized.
Furthermore, this ranking-weighted ensemble approach allows us to make use of models without knowing for which workloads they were originally trained.
Combined with the lightweight nature of BO models, only limited information is required.
These advantages are further illustrated in~\autoref{fig:karasu_intuition}.

\paragraph{Sharing}
We have already noted that in order to train a model $f_i$ for a workload $w_i$ and optimize it with respect to the $k$-th performance measure, only the set of observations $D_i$ is required, which itself evidently contains only limited data and can be derived from all runs with associated data tuples of the form $(w_i, c_j, l_{ij}, y_{ij})$.
Since $w_i$ is not explicitly included in $f_i$ and also otherwise not used, we need a unique identifier $z_i$ to sufficiently group executed profiling runs and their associated data in a repository, which can be realized by any suitable ID generator function.
We from here on use $w_i$ and $z_i$ as well as $W$ and $Z$ interchangeably. 
Finally, to prevent rapid growth of the repository by collaborators uploading excessive metric data $l_{ij}$, we seek a suitable aggregation function $agg: \mathbb{R}^{n \times t} \rightarrow \mathbb{R}^{n \times b}$ that significantly reduces the amount of data, i.e. $b \ll t$, while at best capturing key characteristics. 
We obtain the aggregated metric matrix with $\tilde{l}_{ij} = agg(l_{ij})$.
Ultimately, this yields the tuple $(z_i, c_j, \tilde{l}_{ij}, y_{ij})$ for any executed run, which is sufficient for modeling with BO, contains only limited information and is hence compact in size, and thus qualifies for sharing in a collaborative setting with other collaborators.
An exemplary representation of most of these information is illustrated in~\autoref{fig:overview}.

\subsection{Selecting Data from a Shared Repository}
\label{sec:approach_selecting}

Of course, a shared repository such as the one described above will, over time, contain data from different types of workloads, which requires a feasible approach to data selection.
We, therefore, follow a scalable and opportunistic approach to selecting a promising subset of data and thereby fitted models for our RGPE routine.
This is done based on our compact metric vectors and a suitable similarity function DIST, since we assume that workloads with presumably similar resource requirements might prefer related resource configurations. 
The compact metric vectors can and should, for instance, encompass utilization values for memory, CPU, network, disk, and other mandatory ones, so that a holistic comparison is possible.
For DIST, we make the assumption that resource configurations that are compared with each other include the same machine type (otherwise a default similarity score is assigned).
For all pairs of resource configurations between two workloads, we compute a similarity score via DIST, determine a scaling factor based on the logarithmic distance between the respective machine counts, and calculate the inverse of the scaling factor.
All gathered scaling factors are then normalized, and eventually, a weighted similarity score can be computed.
The intuition behind this is that for many metrics, it holds true that with increasing or decreasing machine count, the utilization of single metrics on individual machines scales proportionally, however, this approximation becomes increasingly inaccurate with diverging machine counts. 
By applying the aforementioned procedure, we obtain an indication of the similarity of two workloads. 
Eventually, candidate workloads are sorted by the obtained similarity scores, and the set of best candidates is returned, with the size of the set depending on the number of support models desired.
The complete procedure is exemplified in~\autoref{alg:similarity_procedure}, for the sake of simplicity in a clearly non-optimized form.
The following helper functions are used:
\begin{itemize}
    \item $runs$: Retrieve all runs (each run is a tuple of data) for a specific workload.
    \item $metrics$: For a single run, get its execution metrics.
    \item $nodes$: For a single run, get the number of machines.
    \item $machineEq$: Determine if two runs were deployed on the same machine type.
\end{itemize}
In addition, we use the Pearson correlation coefficient within our similarity measure function (DIST) because it provides an indication of positive, negative, or no correlation between two workloads and their resource usage profiles, thus eliminating mismatch candidates at first glance.

\begin{algorithm}
\small
\caption{Procedure of Similarity-Based Data Selection}
\label{alg:similarity_procedure}
\begin{algorithmic}
\Require $z_i, Z$
\State $\ $
\Function{dist}{$r_n, r_m$} \Comment{used as similarity measure}
\State $weight \gets abs(log_2(nodes(r_n)) - log_2(nodes(r_m)))$
\State $score \gets pearsonr(metrics(r_n), metrics(r_m))$
\State \Return $1 / (2^{weight})$, $(score + 1) / 2$
\EndFunction
\State $\ $
\State $results \gets new\ List()$
\ForAll{$z_j \in Z: z_i \ne z_j$} \Comment{iterate over candidates}
\State $distResults \gets new\ List()$
\ForAll{$r_n \in runs(z_i)$} \Comment{iterate over run-tuples}
\ForAll{$r_m \in runs(z_j): machineEq(r_n, r_m)$}
\State $distResults.append(\Call{dist}{r_n, r_m})$
\EndFor
\EndFor
\State $results.append((z_j, \text{WeightedAverage}(distResults))$
\EndFor
\State $results \gets sorted(results)$ \Comment{sort candidates by feasibility}
\State \Return $bestResults \subseteq results$ \Comment{return best candidates}
\end{algorithmic}
\end{algorithm}

The proposed selection strategy in conjunction with RGPE has two major advantages.
First, our selection strategy attempts to identify and select data, and thus fitted models, from workloads with presumably similar resource requirements. 
It also indirectly favors the selection of workloads that have been run more frequently, since they are less likely to require the potentially imprecise operation of the logarithmic transformation mentioned above.
On the other hand, even in the case of unpromising selections, RGPE's ranking loss, as well as additional internal mechanisms to prevent weight dilution, allow for the restriction or exclusion of individual models.
This allows our approach to cope with workloads that have different amounts of data stored in a shared repository.
Eventually, candidates are selected based on the re-computed scores.

\subsection{Optimizing for Multiple Objectives}
\label{sec:approach_optimizing}

Related works on resource configuration profiling are primarily concerned with cost reduction considering formulated runtime constraints.
However, in certain situations, both users and service providers may wish to optimize with respect to multiple objectives.
A prominent example is the reduction of energy consumption and thus the carbon footprint of workloads.
As carbon pricing mechanisms, such as emissions trading schemes or carbon taxes, begin to be implemented around the world~\cite{WorldBank_CarbonPricing_2020}, users of cloud services will likely soon be required to report on the emissions associated with their use of cloud infrastructure. This is evidenced by new carbon reporting features being implemented by major cloud providers such as AWS or Google, as well as recent research efforts to transparently quantify the carbon footprint of workloads~\cite{Eilam_CloudCarbonAccounting_2021}.

To account for this, the modeling approach motivated in~\autoref{sec:approach_sharing} can be easily extended to support multiple objectives. 
First, each objective and each constraint is modeled individually using a separate BO model (Gaussian Process), which allows for the application of RGPE for each.
In this context, the expected improvement of one or more objectives is weighted by the probability of feasibility under one or more modeled constraints.
Next, by explicitly treating the individual RGPE models as independent, we can simply optimize the sum of their marginal log likelihoods. 
While in theory this may leave potential untapped in the case of correlated objectives, in practice it makes our approach directly applicable to various problem scenarios without prior incorporation of correlation information.
Conveniently, our data selection strategy remains unaffected by this modeling extension, since it relies only on computed and dense metric information.
Moreover, this decoupled approach allows us to handle workloads that have been optimized with different objectives and/or constraints, since individual models can be dynamically included or excluded.

\section{Evaluation}
\label{sec:evaluation}
This section presents our prototype implementation, the utilized dataset, and our experiments together with a discussion of the results. 
All specifications, software versions, and experiment artifacts are provided in our repository noted in~\autoref{sec:introduction}.

\subsection{Dataset and Preparation}
\label{sec:eval_dataset_preparation}
We evaluate Karasu on a large, publicly available dataset\footnote{\url{https://github.com/oxhead/scout}, Accessed: July 2023} that has previously been used to profile state-of-the-art optimizers~\cite{HsuNFM18,WillTBSK22ruya,hsu2018micky} for resource configurations in public clouds.
It includes data obtained from 18 workloads executed on 69 different configurations (scaleout, VM type) on AWS infrastructure in a multi-node setting (one run per configuration), yielding 1242 total workload executions.
Its prior usage and consistency with our defined assumptions make this dataset ideal for our evaluation.
Each included workload operates on a given data input and/or parametrization using one of many algorithms (extracted from HiBench\footnote{\url{https://github.com/Intel-bigdata/HiBench}, accessed: July 2023} and spark-perf\footnote{\url{https://github.com/databricks/spark-perf}, accessed: July 2023}), like PageRank or Naive-Bayes, implemented on either Hadoop 2.7, Spark 2.1, or Spark 1.5.

For each of the 1242 runs performed, general information such as workload completion and runtime are reported, as well as low-level performance metrics that are exported for each cluster node every 5 seconds via the \texttt{sar} command, which introduces a fairly negligible data collection overhead.
Additionally, we derive general VM type specifications as well as the cost for each run using the prices\footnote{\url{https://instances.vantage.sh}, Accessed: July 2023} for AWS on-demand instances in the USA East Northern Virginia region.
Lastly, to evaluate our approach with respect to multi-objective optimization, we also derive the associated energy consumption of each run by using the Teads Engineering carbon footprint estimator for AWS instances\footnote{\url{https://engineering.teads.com/sustainability}, Accessed: July 2023}. 
In particular, we derive the energy consumption using a linear power profile that depends only on CPU consumption, where the energy consumption at CPU idle and at CPU full load denote the boundaries.
The \emph{xlarge} and \emph{2xlarge} sizes are associated with two and four times the power consumption of the \emph{large} size, respectively.

\subsection{Baselines, Prototype Implementation and Setup}
\label{sec:eval_baselines}

We select two peer-reviewed methods from the related work as baselines. 
Both baselines and Karasu are implemented in the BoTorch~\cite{BalandatKJDLWB20} framework which is based on PyTorch.

\begin{itemize}
    \item \textbf{Naive BO}: We use Bayesian Optimization as suggested in \emph{CherryPick}~\cite{AlipourfardLCVY17}.
    We use Gaussian Process with Matern5/2 kernel as prior function, and Expected Improvement (EI) as the acquisition function. 
    Since our approach makes use of RGPE, which was designed to work with Gaussian processes, this baseline is of particular interest to us and will be our primary focus.
    
    \item \textbf{Augmented BO}: We also follow the idea of Augmented Bayesian Optimization described in \emph{Arrow}~\cite{HsuNFM18}. 
    Here, the Extra-Trees algorithm is used as prior function, and we also use EI as an acquisition function because the actual acquisition function is not well described in the original paper and the corresponding authors did not respond to our request for more information.
    Moreover, the average values of six low-level performance metrics are incorporated into the prior function. 
    
\end{itemize}

Our baselines adhere to the following configurations.
Initial models are obtained by training on three randomly selected samples. 
Note that both methods define their search space differently due to their employed resource configuration encoding approaches.
Also, we assume normally distributed noise with $\mu=0$ and $\sigma^2 = 0.1$ inherent to all observations.
Each objective and each constraint is modeled individually using a separate instance of the respective prior function.
The BoTorch acquisition functions used for single-objective optimization and multi-objective optimization are based on Monte Carlo (MC) sampling and can work with noisy observations.

For the prior function of AugmentedBO, we use the Extra-Trees implementation from \emph{scikit-learn} and adopt most of the default parameters.
The trained model is converted to a PyTorch model via the \emph{Hummingbird} library for faster inference when sampling from the posterior. 
We follow the instructions in the original paper for updating the prior function.
For Karasu, we employ RGPE~\cite{feurer2022practical} for each objective and each constraint, with implemented mechanisms to prevent weight dilution.
After each new observation for a target workload, we execute~\autoref{alg:similarity_procedure} to select the desired amount of most similar data, thus ensuring that at any point in time, we make the optimal selection according to available knowledge. 
As for metrics to collect during execution and to utilize within our selection strategy, we choose \textcolor{\myvariablecolor}{six} percentage metrics exported by \emph{sar} which are often selected in research and system analysis, namely \emph{cpu.\%idle}, \emph{memory.\%memused}, \emph{disk.\%util}, \emph{network.\%ifutil}, \emph{swap.\%swpused}, and \emph{paging.\%vmeff}, and compute for each metric \textcolor{\myvariablecolor}{three} quantile values (10th, 50th, 90th) across time and employed machines, resulting in a compact metric vector, with \textcolor{\myvariablecolor}{three} values for each metric.
The metric vector is then later used in the context of~\autoref{alg:similarity_procedure}.

\subsection{Experiment Design}
\label{sec:eval_experiment_design}
We emulate a shared performance data repository, which requires appropriate data generation using our baselines.
For each workload, we examine \textcolor{\myvariablecolor}{five} different equally spaced percentiles which are used to artificially determine a runtime target among the corresponding run resource configurations. 
Using the subset of workload executions adhering to the runtime target, we can then determine the globally optimal value for each individual objective.
Next, the respective optimization method is trying to find a good resource configuration under consideration of the runtime targets.
This optimization procedure is conducted \textcolor{\myvariablecolor}{ten} times to account for randomness in the initialization phase. 
In total, we conduct \textcolor{\myvariablecolor}{50} unique experiments per workload, each consisting of at most \textcolor{\myvariablecolor}{20} profiling runs -- a reasonable upper bound in light of our search space, deduced from related works~\cite{AlipourfardLCVY17,HsuNFM18}. 
This is repeated for every baseline. 
The generated data is used in the following sections both to visualize the capabilities of individual baselines and to provide Karasu with a data source. 

\subsection{Results}
We evaluate our approach using different scenarios and performance indicators. 
For each scenario to come, we run experiments for all workloads and the aforementioned percentiles for the runtime target determination over \textcolor{\myvariablecolor}{five} iterations.

\textbf{General Performance Boost.} What can be gained from our proposed solution, and how many support models does it take for the observed performance gain to converge?
We evaluate a scenario where the available support models come from the same workload, but are initialized differently and trained with different runtime targets.
This describes a near-optimal situation, as the models used retain valuable information that can be applied almost directly. 
In addition, we investigate the benefit of an increasing number of models.
We do not consider~\autoref{alg:similarity_procedure}, but randomly select the required number of models from all remaining candidate models after filtering.

\begin{figure}[t!]
    \centering
    \includegraphics[width=1\columnwidth]{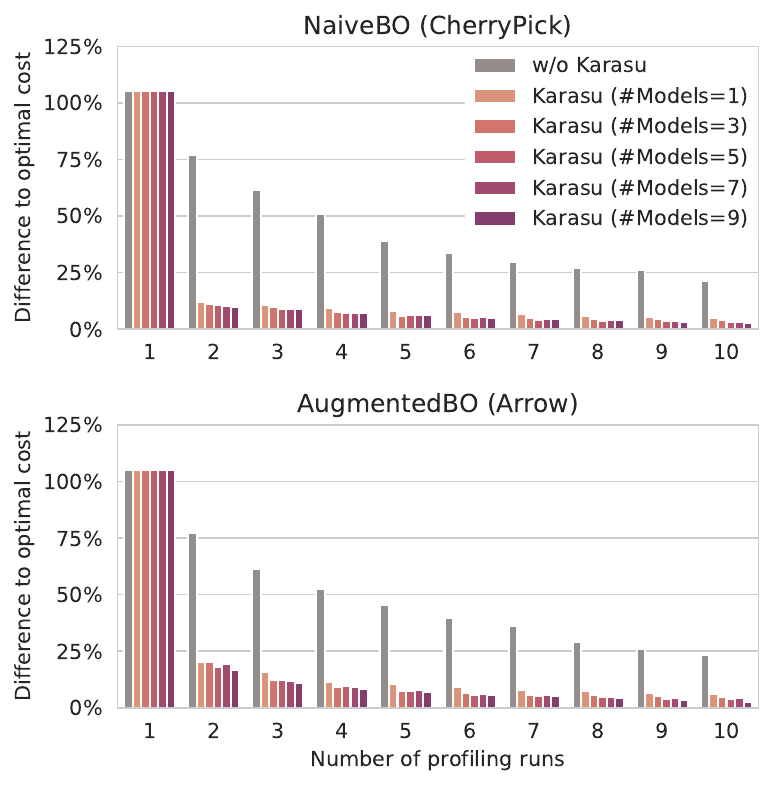}
	\caption{Illustration of the least expensive configuration found under constraints. In general, Karasu significantly accelerates the profiling procedure.
	}
    \label{fig:rq1}
\end{figure}

\autoref{fig:rq1} illustrates the results in terms of cost by showing the cheapest valid resource configuration found so far for the baselines and different Karasu configurations.
The results are identical for the first profiling run because Karasu requires an initial execution to leverage its workload information.
Evidently, Karasu significantly accelerates the profiling search and reduces the cold start problem:
For NaiveBO, the difference to the optimal cost can be reduced to less than 25\,\% already at the \textcolor{\myvariablecolor}{second} profiling run in \textcolor{\myvariablecolor}{88.4\,\%} to \textcolor{\myvariablecolor}{90.2\,\%} of all cases using Karasu, in contrast to \textcolor{\myvariablecolor}{33.0\,\%} for NaiveBO w/o Karasu.
After the \textcolor{\myvariablecolor}{fifth} profiling run, \textcolor{\myvariablecolor}{21.4\,\%} to \textcolor{\myvariablecolor}{26.3\,\%} of all cases using Karasu detected the optimal configuration, in contrast to only \textcolor{\myvariablecolor}{5.8\,\%} for NaiveBO w/o Karasu.
Both baselines not only require significantly more profiling runs to find near-optimal configurations, but also appear to be more sensitive to different runtime targets and initial starting points.
With respect to the number of models, we observe that the performance gains from using more models are quite small in this scenario, since a single available model already speeds up the profiling procedure by a significant margin.
This is in favor of our collaborative approach and makes it applicable even when only small-sized performance data repositories are available.

\begin{figure}[t!]
    \centering
    \includegraphics[width=1\columnwidth]{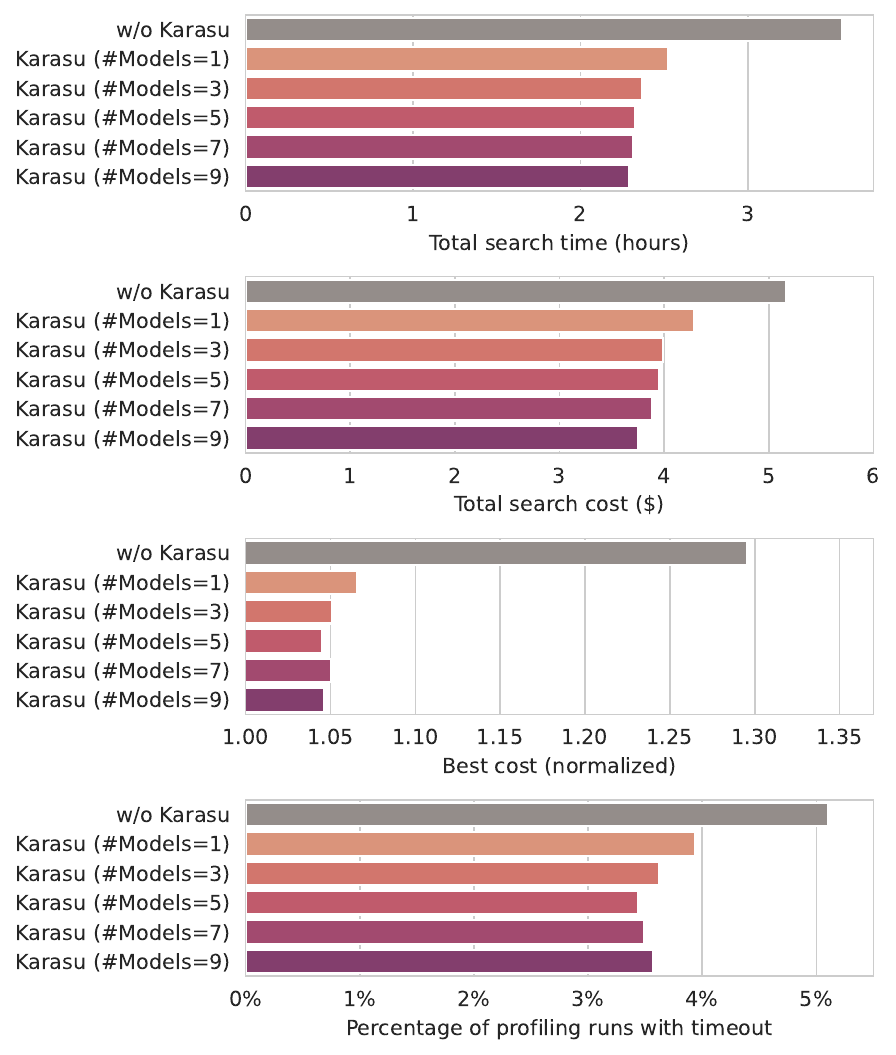}
	\caption{NaiveBO with Karasu: Multiple models reduce search time, search cost, and also tend to find more cost-efficient resource configurations. 
	}
    \label{fig:rq1_total_cost}
\end{figure}

Related works commonly formulate a stopping condition to terminate the profiling after a reasonable number of runs to reduce overhead.
We apply the same procedure as in NaiveBO (CherryPick) and report our findings in~\autoref{fig:rq1_total_cost}: The profiling is stopped when the EI acquisition value of the next most promising candidate satisfies $<=10\,\%$ and at least \textcolor{\myvariablecolor}{six} profiling runs have been conducted.
It can be observed that Karasu achieves superior results across all its model number configurations. 
In addition, it appears that more models reduce total search time and cost while recommending more cost-effective resource configurations, while Karasu generally suggests fewer resource configurations that lead to execution timeouts. 
This is reasonable because the models included help to capture a holistic understanding of the search space.

\textbf{Collaborative Applicability.} How well does Karasu perform in a collaborative scenario, with potentially different workloads and limited available data?
We evaluate a scenario where all the data in the repository comes from different workloads, with individual characteristics, resource requirements, and constraints.
This describes a realistic situation, especially for an individualized target workload.
We set the number of models to \textcolor{\myvariablecolor}{three}, employ~\autoref{alg:similarity_procedure}, and evaluate its applicability by filtering the data in the repository according to different data availability cases:

\begin{itemize}
    \item \emph{Case A}: different framework, algorithm \& dataset
    \item \emph{Case B}: same framework; different algorithm \& dataset
    \item \emph{Case C}: same framework \& algorithm; different dataset
    \item \emph{Case D}: same framework, algorithm \& dataset
\end{itemize}
By gradually forcing previously executed workloads to be dissimilar to the target workload, we eliminate the likelihood of random good choices and allow for detailed evaluation.
In other words, we investigate decreasingly restrictive scenarios.

\begin{figure}[t!]
    \centering
    \includegraphics[width=1\columnwidth]{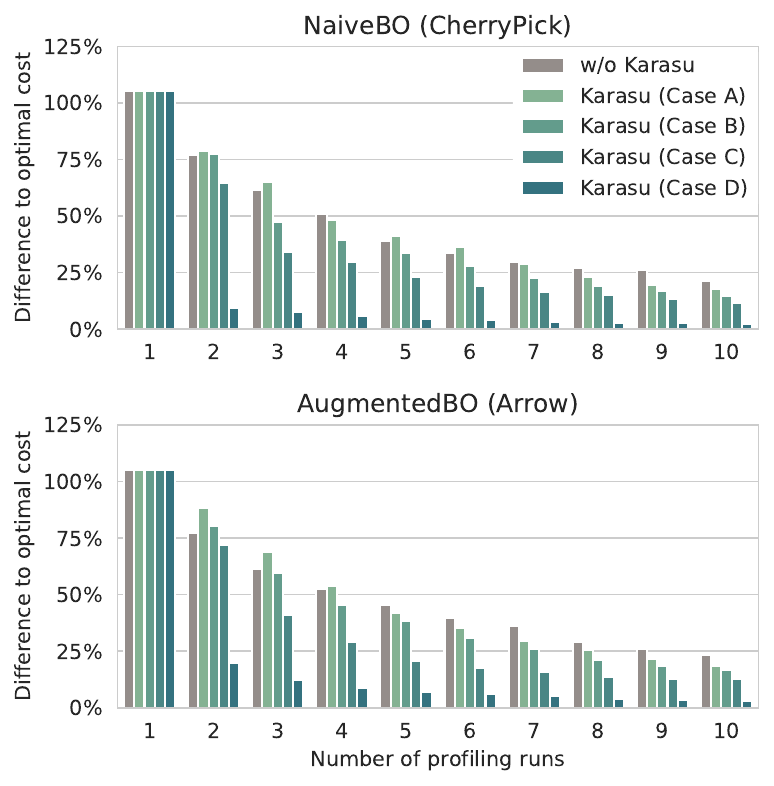}
	\caption{Illustration of the least expensive configuration found under constraints. Data availability evidently impacts upon the chance for improvement.
	}
    \label{fig:rq2}
\end{figure}

\autoref{fig:rq2} illustrates the results in terms of cost.
Evidently, the performance improvements depend on the data availability. 
While significant improvements can be observed for Case C and especially Case D, no relevant improvement can be achieved for Case A.
This is to be expected, as the data availability is very limited in this scenario: different frameworks may follow opposing computing paradigms, so the information gathered is not easily transferable.
However, it is worth noting that in such a scenario, Karasu quickly recognizes this and puts less emphasis on the selected support models, which is manifested in the comparable convergence behavior between the baseline (w/o Karasu) and Karasu (Case A).
An additional general factor for performance improvements is both the diversity of data in the shared repository as well as the scope of the resource configuration search space.
In another line of experiments on a dataset with significantly more workloads but smaller search spaces, we find that performance improvements are strongly pronounced across all four data availability cases.
We observe that, depending on the data availability case, Karasu allows for equal or better performance than the respective baselines, demonstrating that it can be used in a collaborative scenario, even with collaborators who do not necessarily use related frameworks, algorithms, or datasets.

\begin{figure}[t!]
    \centering
    \includegraphics[width=1\columnwidth]{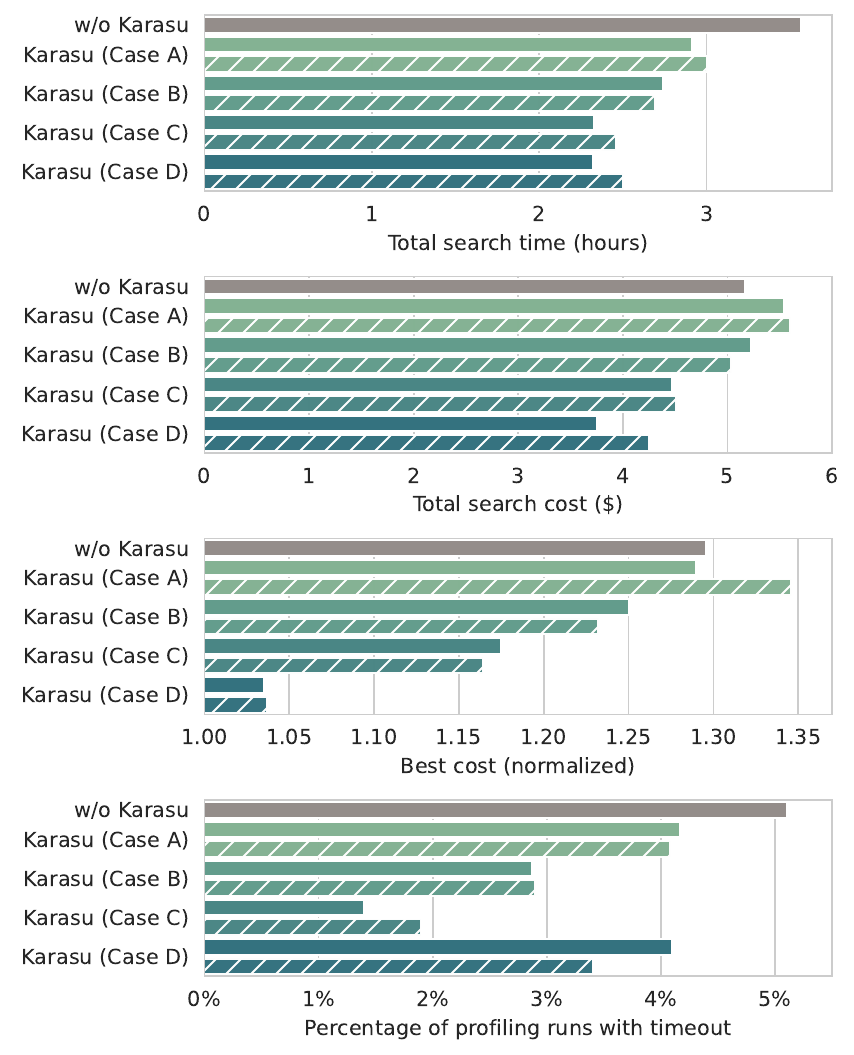}
	\caption{
	NaiveBO with Karasu: Improved results can be achieved across relevant performance indicators depending on the data availability case.
    Hatched bars indicate data availability cases with heterogeneous data amounts.
	}
    \label{fig:rq2_total_cost}
\end{figure}

As before, we evaluate the early stopping scenario and report our results in~\autoref{fig:rq2_total_cost} using the unhatched bars.
It is evident that depending on the concrete data availability case, Karasu outperforms the employed baselines across most or all performance indicators.
While performance degradations can be observed as data availability is increasingly restricted, Karasu remains advantageous and allows for finding near-optimal resource configurations while reducing overall search time and cost. 
It is also worth mentioning that in the case of very homogeneous data sources such as Case D, timeouts are less likely to be prevented, due to the lack of diverse information in the employed support models. 

We also evaluate the effect of heterogeneous data amounts on the capabilities of our approach.
To emulate this, we modify the available shared data for each target workload, i.e., we consider only the first $3 \leq k \leq n$ profiled data points of each candidate workload, with $k$ being randomly sampled assuming a uniform distribution, and $n$ denoting the total number of profiled data points. 
Evidently, this operation affects our data selection procedure and the goodness of individually trained models. 
The results are reported in~\autoref{fig:rq2_total_cost} using the hatched bars.
It can be observed that although heterogeneous data seems to affect our approach, the impact is fairly negligible and not always unambiguous.
Notably, heterogeneous data tend to slightly increase the overall search time and cost, which leads to finding slightly more cost-optimal resource configurations in most cases.
On the other hand, the effects on the probability of choosing more timeout-prone resource configurations are not unambiguous.
We conclude that for most investigated cases, the advantages of using Karasu remain significant, as undesired effects associated with heterogeneous shared data are regulated and limited by our data selection procedure and model weighting mechanisms. 

\begin{figure}
    \centering
    \includegraphics[width=.75\columnwidth]{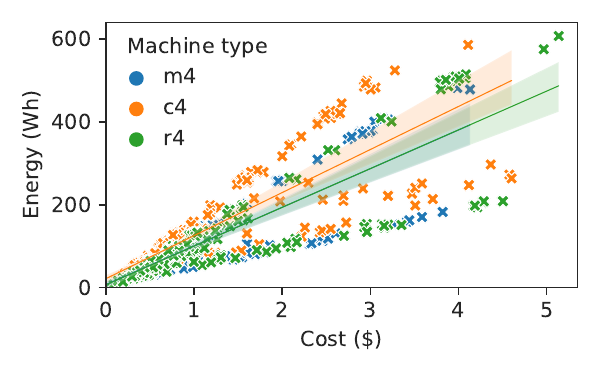}
    \caption{Energy consumption and cost of all workloads in our evaluation dataset, grouped by machine type and accompanied by a regression estimate.}
    \label{fig:objective_correlations}
\end{figure}

\textbf{Multi-Objective Support.} To evaluate Karasu for multi-objective optimization (MOO), we additionally use energy consumption as a second objective. 
It can be directly derived from the performance metrics and is linearly correlated with the associated carbon emissions. 
\autoref{fig:objective_correlations} shows the energy consumption and cost of all workloads in the dataset.
We can see that the two objectives are correlated, especially when approaching their minimum.
Therefore, we will show that (i) we can achieve lower energy consumption faster with MMO while accepting slightly higher costs and (ii) that Karasu is heavily improving the results compared to the baseline NaiveBO approach.
We consider \textcolor{\myvariablecolor}{Case~D} with \textcolor{\myvariablecolor}{three} models.

\begin{figure}
    \centering
    \includegraphics[width=\columnwidth]{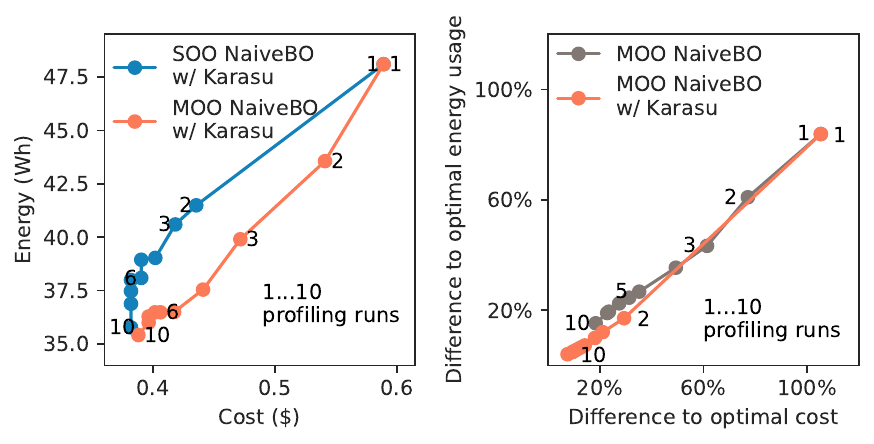}
    \hspace{0.2cm}
    \begin{minipage}[t]{.45\columnwidth}
		\caption{Example of single- vs. multi-objective optimization using NaiveBO with Karasu.}\label{fig:rq3_cherrypick}
	\end{minipage}
    \hspace{0.3cm}
	\begin{minipage}[t]{.45\columnwidth}
		\caption{Average case results of multi-objective optimization for NaiveBO with and without Karasu.}\label{fig:rq3}
	\end{minipage}
\end{figure}

\autoref{fig:rq3_cherrypick} presents an example of when multi-objective optimization is preferable over single-objective optimization.
While SOO initially explores inexpensive resource configurations for the price of higher energy consumption, MOO balances both objectives fairly early and eventually finds a slightly more expensive, but more energy-efficient resource configuration. 
This shows that MOO can be generally useful for resource configuration profiling, especially for independent objectives.
Furthermore, we investigated how Karasu can accelerate profiling in a MOO setting.
The results are depicted in~\autoref{fig:rq3}, where we report the average results for NaiveBO, with and without Karasu.
Evidently, Karasu allows for quick initial progress and then converges towards an optimum, whereas the baseline can not rely on transferred knowledge and thus requires more profiling for similar results.
This confirms our previous findings and underscores the performance gain that Karasu can bring to existing approaches, regardless of single-objective or multi-objective optimization.

\subsection{Discussion}

In light of our previous findings, we will in the following discuss the applicability and overhead of our approach Karasu.

\textbf{Applicability.}
With our various conducted experiments, we demonstrated the potential of Karasu to accelerate resource configuration profiling, in light of different numbers of support models, data availability cases, and optimization objectives. 
Even a few support models are capable of significantly boosting the profiling performance. Furthermore, Karasu is well suited for use in a collaborative setting, as it allows the reuse of existing performance models when workloads change~\cite{AlipourfardLCVY17} and leverages the similarity in algorithm implementations~\cite{will2021c3o} or reliance on established data processing frameworks~\cite{0001DBTGHMH19}. Karasu also abstracts away the selection and consideration of promising support models, promising equal or better performance than related methods based on our results.

\textbf{Overhead.}
The overhead of using Karasu can be attributed to two main aspects. Regarding \underline{sharing}, the most time-consuming task is the selection procedure. 
In a real-world scenario with a shared repository and underlying database, this can be efficiently realized through proper indexing and a respective distance operator. 
Individual models can also be saved next to the associated data, reducing data transmission and computational effort on the client side. 
Karasu is designed to save data by considering aggregated resource metrics, performance measures, and common knowledge about the resource configuration, without using the workload of interest.%

For \underline{modeling}, the time required for training and prediction depends on the number of data samples and models. 
Since Karasu operates on limited data samples, the time required for model training can be neglected. 
The training of support models can be outsourced to the repository, requiring only the local ranking of models, or the training of individual models can be parallelized before ranking. 
Another source of overhead is the inference phase, due to Monte Carlo sampling in RGPE. 
However, the observed overhead is in the range of seconds, which aligns with the formal complexity of RGPE~\cite{feurer2022practical}.

\section{Related Work}
\label{sec:related_work}

We classify existing resource configuration works into \textit{Iterative Search-Based} and \textit{Performance Model-Based} approaches.

\subsection{Iterative Search-Based Approaches}

Without knowledge about prior workload executions, profiling is required to determine appropriate resource configurations.
Approaches have been proposed in the past that primarily focus on the profiling phase and attempt to reduce the associated overhead.
Older works~\cite{AlipourfardLCVY17,HsuNFM18} leverage a form of BO to find near-optimal resource configurations for a single workload, sometimes under consideration of low-level performance metrics.
To reduce the overhead of finding the optimal configuration, a configurable threshold on the model confidence is often used to stop further searches.
Other works extend this approach to resource configuration optimization for a set of workloads~\cite{hsu2018micky} simultaneously, e.g. if a resource configuration shall be used for diverse applications, or explicitly exploit similarities~\cite{bilal2020finding,ScheinertBBTWK22} of workload characteristics, e.g. via benchmarking runs, to make the search process more effective.
Another strategy proposed in some works~\cite{klimovic2018selecta,WillTBSK22ruya,AlSayehMJPS22,AlSayehJMS22} includes lightweight workload profiling to then better navigate the search space efficiently.%

More recent works~\cite{fekry2020accelerating,FekryCPRH20} employ multitask BO or other ensemble methods to leverage information from executions of comparable workloads in order to find near-optimal configurations faster.
Karasu can be considered closest to those approaches, but explicitly assumes a collaborative scenario for distributed dataflows, focusing on minimal data sharing and collaborative applicability.
Due to its design, Karasu also qualifies for integration with existing BO-based methods.

\subsection{Performance Model-Based Approaches}

If sufficient knowledge about prior runs is available, more comprehensive performance models can be trained to yield resource configuration recommendations, e.g. with respect to defined runtime targets.
A prominent example is Ernest~\cite{venkataraman2016ernest}, a performance prediction framework for large-scale advanced analytics workloads that leverages parametric modeling and statistical techniques.
Its limitations have been addressed in works such as~\cite{RajanKCK16,AlSayehMJPS22}, e.g. by automatically choosing between either parametric or non-parametric models to optimally support given workloads, or by automatically selecting appropriate datasets to cache.
Other works~\cite{YadwadkarHGSK17,ScheinertTZWAWK21,ChenLLWZ21silhouette} support using historical performance data, possibly from executions on different infrastructures, to apply transfer learning on performance models for a given job.

As access to own historical data is often a limiting factor or strong assumption, our prior works~\cite{will2021c3o,ScheinertTZWAWK21} motivate collaborative approaches which leverage data sharing among collaborators that employ similar processing frameworks or algorithm implementations.
This way, users can benefit from existing performance models already at the first execution of a job in a new context, at the cost of sharing detailed and potentially sensitive information with others.
Building upon that, Karasu additionally focuses on minimal data sharing and lightweight, easily composable performance models.
\section{Conclusion}
\label{sec:conclusion}
We present Karasu, a collaborative approach for improving resource configuration profiling of compute clusters running big data analytics jobs.
Karasu promotes data sharing among collaborators that operate similar workloads to build lightweight performance models that enable more efficient profiling with respect to user-defined objectives and constraints.
Our approach uses an ensemble method and combines it with a data selection strategy that relies on the similarity of compact vectors representing the resource requirements of individual workloads.
We show that Karasu finds near-optimal configurations significantly faster and at a lower cost than existing approaches, even when only a few comparable models are available.
In the future, we plan to investigate how to benchmark arbitrary infrastructures in a representative manner to further dissect the search space and accelerate the search.

\section*{Acknowledgments}
This work has been supported through a grant by the German
Research Foundation (DFG) as C5  (grant 506529034).

\bibliographystyle{IEEEtran}
\bibliography{bib}

\begin{thebibliography}{10}
\providecommand{\url}[1]{#1}
\csname url@samestyle\endcsname
\providecommand{\newblock}{\relax}
\providecommand{\bibinfo}[2]{#2}
\providecommand{\BIBentrySTDinterwordspacing}{\spaceskip=0pt\relax}
\providecommand{\BIBentryALTinterwordstretchfactor}{4}
\providecommand{\BIBentryALTinterwordspacing}{\spaceskip=\fontdimen2\font plus
\BIBentryALTinterwordstretchfactor\fontdimen3\font minus
  \fontdimen4\font\relax}
\providecommand{\BIBforeignlanguage}[2]{{%
\expandafter\ifx\csname l@#1\endcsname\relax
\typeout{** WARNING: IEEEtran.bst: No hyphenation pattern has been}%
\typeout{** loaded for the language `#1'. Using the pattern for}%
\typeout{** the default language instead.}%
\else
\language=\csname l@#1\endcsname
\fi
#2}}
\providecommand{\BIBdecl}{\relax}
\BIBdecl

\bibitem{Google2020Autopilot}
K.~Rzadca, P.~Findeisen, J.~Swiderski, P.~Zych, P.~Broniek, J.~Kusmierek,
  P.~Nowak, B.~Strack, P.~Witusowski, S.~Hand, and J.~Wilkes, ``Autopilot:
  workload autoscaling at google,'' in \emph{EuroSys}.\hskip 1em plus 0.5em
  minus 0.4em\relax {ACM}, 2020.

\bibitem{cloudcomputingchallenges2018}
Y.~Al-Dhuraibi, F.~Paraiso, N.~Djarallah, and P.~Merle, ``Elasticity in cloud
  computing: State of the art and research challenges,'' \emph{IEEE
  Transactions on Services Computing}, vol.~11, no.~2, 2018.

\bibitem{RajanKCK16}
K.~Rajan, D.~Kakadia, C.~Curino, and S.~Krishnan, ``Perforator: eloquent
  performance models for resource optimization,'' in \emph{SoCC}.\hskip 1em
  plus 0.5em minus 0.4em\relax {ACM}, 2016.

\bibitem{venkataraman2016ernest}
S.~Venkataraman, Z.~Yang, M.~J. Franklin, B.~Recht, and I.~Stoica, ``Ernest:
  Efficient performance prediction for large-scale advanced analytics,'' in
  \emph{NSDI}.\hskip 1em plus 0.5em minus 0.4em\relax {USENIX}, 2016.

\bibitem{ShahAKW19}
S.~Shah, Y.~Amannejad, D.~Krishnamurthy, and M.~Wang, ``Quick execution time
  predictions for spark applications,'' in \emph{CNSM}.\hskip 1em plus 0.5em
  minus 0.4em\relax {IEEE}, 2019.

\bibitem{AlSayehS19}
H.~Al{-}Sayeh and K.~Sattler, ``Gray box modeling methodology for runtime
  prediction of apache spark jobs,'' in \emph{ICDE}.\hskip 1em plus 0.5em minus
  0.4em\relax {IEEE}, 2019.

\bibitem{KirchoffXMR19}
D.~F. Kirchoff, M.~G. Xavier, J.~Mastella, and C.~A. F.~D. Rose, ``A
  preliminary study of machine learning workload prediction techniques for
  cloud applications,'' in \emph{PDP}.\hskip 1em plus 0.5em minus 0.4em\relax
  {IEEE}, 2019.

\bibitem{ChenLLWZ21silhouette}
Y.~Chen, L.~Lin, B.~Li, Q.~Wang, and Q.~Zhang, ``Silhouette: Efficient cloud
  configuration exploration for large-scale analytics,'' \emph{{IEEE} Trans.
  Parallel Distributed Syst.}, vol.~32, no.~8, 2021.

\bibitem{AlSayehMJPS22}
H.~Al{-}Sayeh, B.~Memishi, M.~A. Jibril, M.~Paradies, and K.~Sattler,
  ``Juggler: Autonomous cost optimization and performance prediction of big
  data applications,'' in \emph{SIGMOD}.\hskip 1em plus 0.5em minus 0.4em\relax
  {ACM}, 2022.

\bibitem{AlipourfardLCVY17}
O.~Alipourfard, H.~H. Liu, J.~Chen, S.~Venkataraman, M.~Yu, and M.~Zhang,
  ``Cherrypick: Adaptively unearthing the best cloud configurations for big
  data analytics,'' in \emph{NSDI}.\hskip 1em plus 0.5em minus 0.4em\relax
  {USENIX}, 2017.

\bibitem{HsuNFM18}
C.~Hsu, V.~Nair, V.~W. Freeh, and T.~Menzies, ``Arrow: Low-level augmented
  bayesian optimization for finding the best cloud {VM},'' in
  \emph{ICDCS}.\hskip 1em plus 0.5em minus 0.4em\relax {IEEE} Computer Society,
  2018.

\bibitem{hsu2018micky}
C.~Hsu, V.~Nair, T.~Menzies, and V.~W. Freeh, ``Micky: {A} cheaper alternative
  for selecting cloud instances,'' in \emph{CLOUD}.\hskip 1em plus 0.5em minus
  0.4em\relax {IEEE} Computer Society, 2018.

\bibitem{bilal2020finding}
M.~Bilal, M.~Canini, and R.~Rodrigues, ``Finding the right cloud configuration
  for analytics clusters,'' in \emph{SoCC}.\hskip 1em plus 0.5em minus
  0.4em\relax {ACM}, 2020.

\bibitem{klimovic2018selecta}
A.~Klimovic, H.~Litz, and C.~Kozyrakis, ``Selecta: Heterogeneous cloud storage
  configuration for data analytics,'' in \emph{ATC}.\hskip 1em plus 0.5em minus
  0.4em\relax {USENIX}, 2018.

\bibitem{fekry2020accelerating}
A.~Fekry, L.~Carata, T.~Pasquier, and A.~Rice, ``Accelerating the configuration
  tuning of big data analytics with similarity-aware multitask bayesian
  optimization,'' in \emph{BigData}.\hskip 1em plus 0.5em minus 0.4em\relax
  {IEEE}, 2020.

\bibitem{MendesCRG20}
P.~Mendes, M.~Casimiro, P.~Romano, and D.~Garlan, ``Trimtuner: Efficient
  optimization of machine learning jobs in the cloud via sub-sampling,'' in
  \emph{MASCOTS}.\hskip 1em plus 0.5em minus 0.4em\relax {IEEE}, 2020.

\bibitem{LiuXL20}
Y.~Liu, H.~Xu, and W.~C. Lau, ``Accordia: Adaptive cloud configuration
  optimization for recurring data-intensive applications,'' in
  \emph{ICDCS}.\hskip 1em plus 0.5em minus 0.4em\relax {IEEE}, 2020.

\bibitem{SongZLSFDS21}
F.~Song, K.~Zaouk, C.~Lyu, A.~Sinha, Q.~Fan, Y.~Diao, and P.~J. Shenoy,
  ``Spark-based cloud data analytics using multi-objective optimization,'' in
  \emph{ICDE}.\hskip 1em plus 0.5em minus 0.4em\relax {IEEE}, 2021.

\bibitem{CasimiroD0RZG20}
M.~Casimiro, D.~Didona, P.~Romano, L.~E.~T. Rodrigues, W.~Zwaenepoel, and
  D.~Garlan, ``Lynceus: Cost-efficient tuning and provisioning of data analytic
  jobs,'' in \emph{ICDCS}.\hskip 1em plus 0.5em minus 0.4em\relax {IEEE}, 2020.

\bibitem{0007SCR20}
M.~Bilal, M.~Serafini, M.~Canini, and R.~Rodrigues, ``Do the best cloud
  configurations grow on trees? an experimental evaluation of black box
  algorithms for optimizing cloud workloads,'' \emph{Proc. {VLDB} Endow.},
  vol.~13, no.~11, 2020.

\bibitem{FekryCPRH20}
A.~Fekry, L.~Carata, T.~F.~J. Pasquier, A.~Rice, and A.~Hopper, ``To tune or
  not to tune?: In search of optimal configurations for data analytics,'' in
  \emph{SIGKDD}.\hskip 1em plus 0.5em minus 0.4em\relax {ACM}, 2020.

\bibitem{ScheinertTZWAWK21}
D.~Scheinert, L.~Thamsen, H.~Zhu, J.~Will, A.~Acker, T.~Wittkopp, and O.~Kao,
  ``Bellamy: Reusing performance models for distributed dataflow jobs across
  contexts,'' in \emph{CLUSTER}.\hskip 1em plus 0.5em minus 0.4em\relax {IEEE},
  2021.

\bibitem{will2021c3o}
J.~Will, L.~Thamsen, D.~Scheinert, J.~Bader, and O.~Kao, ``{C3O: Collaborative
  Cluster Configuration Optimization for Distributed Data Processing in Public
  Clouds},'' in \emph{IC2E}.\hskip 1em plus 0.5em minus 0.4em\relax IEEE, 2021.

\bibitem{WorldBank_CarbonPricing_2020}
{World Bank}, ``State and trends of carbon pricing 2020,'' Washington, DC:
  World Bank., Tech. Rep., 2020.

\bibitem{Eilam_CloudCarbonAccounting_2021}
T.~Eilam, ``Towards transparent and trustworthy cloud carbon accounting,'' in
  \emph{Extended Abstracts of Middleware '21}.\hskip 1em plus 0.5em minus
  0.4em\relax ACM, 2021.

\bibitem{feurer2022practical}
M.~Feurer, B.~Letham, F.~Hutter, and E.~Bakshy, ``Practical transfer learning
  for bayesian optimization,'' 2022.

\bibitem{WillTBSK22ruya}
J.~Will, L.~Thamsen, J.~Bader, D.~Scheinert, and O.~Kao, ``Ruya: Memory-aware
  iterative optimization of cluster configurations for big data processing,''
  in \emph{BigData}.\hskip 1em plus 0.5em minus 0.4em\relax {IEEE}, 2022.

\bibitem{BalandatKJDLWB20}
M.~Balandat, B.~Karrer, D.~R. Jiang, S.~Daulton, B.~Letham, A.~G. Wilson, and
  E.~Bakshy, ``Botorch: {A} framework for efficient monte-carlo bayesian
  optimization,'' in \emph{NeurIPS}, 2020.

\bibitem{0001DBTGHMH19}
G.~Nguyen, S.~Dlugolinsky, M.~Bob{\'{a}}k, V.~D. Tran, {\'{A}}.~L.
  Garc{\'{\i}}a, I.~Heredia, P.~Mal{\'{\i}}k, and L.~Hluch{\'{y}}, ``Machine
  learning and deep learning frameworks and libraries for large-scale data
  mining: a survey,'' \emph{Artif. Intell. Rev.}, vol.~52, no.~1, 2019.

\bibitem{ScheinertBBTWK22}
D.~Scheinert, S.~Becker, J.~Bader, L.~Thamsen, J.~Will, and O.~Kao, ``Perona:
  Robust infrastructure fingerprinting for resource-efficient big data
  analytics,'' in \emph{BigData}.\hskip 1em plus 0.5em minus 0.4em\relax
  {IEEE}, 2022.

\bibitem{AlSayehJMS22}
H.~Al{-}Sayeh, M.~A. Jibril, B.~Memishi, and K.~Sattler, ``Blink: Lightweight
  sample runs for cost optimization of big data applications,'' in
  \emph{ADBIS}.\hskip 1em plus 0.5em minus 0.4em\relax Springer, 2022.

\bibitem{YadwadkarHGSK17}
N.~J. Yadwadkar, B.~Hariharan, J.~E. Gonzalez, B.~Smith, and R.~H. Katz,
  ``Selecting the \emph{best} {VM} across multiple public clouds: a data-driven
  performance modeling approach,'' in \emph{SoCC}.\hskip 1em plus 0.5em minus
  0.4em\relax {ACM}, 2017.

\end{thebibliography}

\end{document}